\documentclass[prl,twocolumn,showpacs,preprintnumbers,amsfonts,amsmath,amssymb,superscriptaddress]{revtex4}
\usepackage{txfonts}
\usepackage{graphicx}
\usepackage{dcolumn}
\usepackage{bm}
\usepackage{array}

\begin{document}

\title{Tuning the distance to a possible ferromagnetic quantum critical point in A$_2$Cr$_3$As$_3$}
\author{J. Luo}
\affiliation{Institute of Physics, Chinese Academy of Sciences,\\
 and Beijing National Laboratory for Condensed Matter Physics,Beijing 100190, China}
\affiliation{School of Physical Sciences, University of Chinese Academy of Sciences, Beijing 100190, China}

\author{J. Yang}
\affiliation{Institute of Physics, Chinese Academy of Sciences,\\
 and Beijing National Laboratory for Condensed Matter Physics,Beijing 100190, China}

\author{R. Zhou}
\affiliation{Institute of Physics, Chinese Academy of Sciences,\\
	and Beijing National Laboratory for Condensed Matter Physics,Beijing 100190, China}

\author{Q. G. Mu}
\affiliation{Institute of Physics, Chinese Academy of Sciences,\\
 and Beijing National Laboratory for Condensed Matter Physics,Beijing 100190, China}
\affiliation{School of Physical Sciences, University of Chinese Academy of Sciences, Beijing 100190, China}

\author{T. Liu}
\affiliation{Institute of Physics, Chinese Academy of Sciences,\\
	and Beijing National Laboratory for Condensed Matter Physics,Beijing 100190, China}
\affiliation{School of Physical Sciences, University of Chinese Academy of Sciences, Beijing 100190, China}

\author{Zhi-an Ren}
\affiliation{Institute of Physics, Chinese Academy of Sciences,\\
 and Beijing National Laboratory for Condensed Matter Physics,Beijing 100190, China}

\author{C. J. Yi}
\affiliation{Institute of Physics, Chinese Academy of Sciences,\\
	and Beijing National Laboratory for Condensed Matter Physics,Beijing 100190, China}
\affiliation{School of Physical Sciences, University of Chinese Academy of Sciences, Beijing 100190, China}

 \author{Y. G. Shi}
\affiliation{Institute of Physics, Chinese Academy of Sciences,\\
	and Beijing National Laboratory for Condensed Matter Physics,Beijing 100190, China}

\author{Guo-qing Zheng}
\thanks{gqzheng123@gmail.com}
\affiliation{Institute of Physics, Chinese Academy of Sciences,\\
	and Beijing National Laboratory for Condensed Matter Physics,Beijing 100190, China}
\affiliation{Department of Physics, Okayama University, Okayama 700-8530, Japan}
\date{\today}

\begin{abstract}
{
	Although superconductivity in the vicinity of antiferromagnetic (AFM) instability has been extensively explored 
	in the last three decades or so, superconductivity in compounds with a background of ferromagnetic (FM) spin fluctuations is still rare.
	We report $^{75}$As nuclear quadrupole resonance  measurements on the 
	A$_2$Cr$_3$As$_3$ family, which is  the first group of  Cr-based superconductors at ambient pressure, with A being alkali elements. 
	From the temperature dependence of the spin-lattice relaxation rate (1/$T_1$),
	we find that by changing A in the order of A=Na, Na$_{0.75}$K$_{0.25}$, K, and Rb,
	the system is tuned to approach a possible FM quantum critical point (QCP).
	This may be ascribed to
	the Cr2-As2-Cr2 bond angle 
	that decreases towards 90$^\circ$, which enhances the FM interaction via the Cr2-As2-Cr2 path.
	 Upon moving away from the QCP, the superconducting transition temperature $T_{\rm sc}$ increases progressively up to 8.0 K in Na$_2$Cr$_3$As$_3$,
	 which is in sharp contrast to the AFM case where $T_{\rm sc}$ usually shows a maximum around  a QCP.
	The 1/$T_1$ decreases rapidly below $T_{\rm sc}$ with no Hebel-Slichter peak, and ubiquitously follows a $T$$^{5}$ variation below a characteristic temperature $T^*$$\approx$0.6 $T_{\rm sc}$, which indicates the existence of point nodes in the superconducting gap function commonly in the 
	 family. These results suggest that the A$_2$Cr$_3$As$_3$ family is a possible solid-state analog of superfluid $^3$He.  }
\end{abstract}

\maketitle
The interplay between magnetism and superconductivity is a key topic in condensed matter physics. In the past thirty years or so, a large amount of superconductors in proximity to an  
 antiferromagnetic (AFM) ordered phase have been found. In heavy fermions\cite{heavyfermion,QCP}, cuprates\cite{cuprate}, 
  and iron pnictides \cite{ironbased} 
  superconductivity appears on the verge of antiferromagnetic instability, and the  critical temperature $T_{\rm sc}$ usually takes a maximum at the  AFM quantum critical point (QCP).  However,  superconductivity in the vicinity of a ferromagnetic ordered phase is still rare. 
UGe$_2$ is a ferromagnet, but becomes superconducting under high pressure when the Curie temperature $T_{\rm C}$ is reduced 
inside the ferromagnetic phase \cite{UGe2}. 
However, superconductivity has not been found when the ferromagnetic order is completely destroyed. 

Ferromagnetic interactions can also promote quantum states other than superconductivity. For example,
superfluidity in $^3$He emerges in the background of ferromagnetic spin fluctuations. 
There are two phases, namely A phase  and B phase, in  $^3$He \cite{He3,ABM1,ABM2,BW}.
The A phase is a  $p$-wave ABM (Anderson-Brinkman-Morel) state with equal spin  pairing ($\uparrow$$\uparrow$ and $\downarrow$$\downarrow$ ) \cite{ABM1,ABM2}. The B phase is a $p$-wave BW (Balian-Werthamer) state with an additional component 1/$\sqrt{2}$($\uparrow$$\downarrow$+$\downarrow$$\uparrow$) \cite{BW}.
 There are point nodes in the  gap function  in ABM 
 state but the gap is isotropic in BW 
  state\cite{leggett}. 
 It has been pointed out that the spin triplet pairing in $^3$He is induced by  FM spin fluctuations \cite{ABM2}.  
 Notably, $^3$He phases are  topologically non-trivial, which has received new and intensive interests  in the past few years \cite{QiandZhang}.  The B phase of $^3$He belongs to the so-called DIII topological class \cite{class}, and the A phase bares similarities to topological Weyl semi-metals. 
 Therefore, searching for a solid state analog of $^3$He serves to bridge three large research areas: strong correlations, unconventional superconductivity, and topological quantum phenomena.

 Recently, a 3$d$-electron system,  chromium-based superconductors A$_2$Cr$_3$As$_3$(A = Na, K, Rb, Cs), has been discovered \cite{Na2Cr3As3,K2Cr3As3,Rb2Cr3As3,Cs2Cr3As3}. In this family, superconductivity emerges from a   paramagnetic state.  K$_2$Cr$_3$As$_3$ and Rb$_2$Cr$_3$As$_3$ have a \emph{T}$_{\rm sc}$=6.1 and  4.8 K \cite{K2Cr3As3,Rb2Cr3As3}, respectively, and  Na$_2$Cr$_3$As$_3$  has the highest \emph{T}$_{\rm sc}$ = 8.0 K. Resistivity  measurement suggests that Cs$_2$Cr$_3$As$_3$ superconducts below $T$=2.2 K \cite{Cs2Cr3As3}. Although anomaly at $T$=2.2 K was not found  by 
 nuclear quadrupole resonance (NQR)  \cite{imaiCsCrAs,Luojununpublished},  magnetic susceptibility measurement does confirm bulk superconductivity below $T$=1.2 K \cite{Sumiyama}.  
 Density function theory (DFT) calculations show that there are three bands across the Fermi level, namely, two quasi one-dimensional (1D) band $\alpha$ and $\beta$, and one three-dimensional (3D) band $\gamma$ \cite{ferromagneticfluctuation1,magnetism}.
 The $\gamma$ band makes the main contribution to the density of states (DOS).

 NQR,   penetration depth,  muon spin rotation ($\mu$SR),  upper critical field \emph{H}$_{\rm c2}$ and specific heat measurements show signatures of unconventional superconductivity \cite{imai,yang,specific,penetration,muSRK2Cr3As3,muSRRb2Cr3As3,Hc2,angularHc2}.
 Theoretically, spin-triplet superconducting state 
 has been proposed \cite{fwave,pwave,spintriplet}. 
  In the normal state, 
  ferromagnetic spin fluctuations have been found from the  Knight shift and spin-lattice relaxation rate (1/\emph{T}$_1$) measurements in Rb$_2$Cr$_3$As$_3$ \cite{yang}.
  Neutron  scattering measurements also suggest short-range magnetic order in K$_2$Cr$_3$As$_3$ \cite{magneticins}. However, how the spin fluctuations evolve with changing A  is unknown. In addition, the gap symmetry is still controversial. For example,  1/\emph{T}$_1$ $\propto$ $T^4$ was reported in K$_2$Cr$_3$As$_3$ \cite{imai}, but 1/\emph{T}$_1$ $\propto$ $T^5$ was found in Rb$_2$Cr$_3$As$_3$ \cite{yang}. The latter result suggests point nodes in the gap function \cite{yang}. On the other hand, line nodes were claimed by penetration depth and specific heat measurements \cite{specific,penetration}. 

In this work, we systematically study the normal  and superconducting states of the  family  A$_2$Cr$_3$As$_3$ 
by NQR. We find that  1/\emph{T}$_1$$T$ in the normal state for all compounds can be fitted by Moriya's theory for 3D FM spin fluctuations  \cite{SCRtheory}. Our results show  that on going from A = Na to Na$_{0.75}$K$_{0.25}$, K, Rb, the system progressively approaches a possible FM QCP, which may be ascribed   to the Cr2-As2-Cr2 bond angle  that decreases toward 90$^\circ$ in the same order and enhances the FM interaction. In the superconducting state, 1/\emph{T}$_1$ for all compounds show no Hebel-Slichter  peak, and 
 follows a \emph{T}$^5$ variation below a characteristic temperature \emph{T}* $\approx$ 0.6\emph{T}$_{\rm sc}$, indicating a common formation of point nodes in the  gap function. 
Our results indicate that A$_2$Cr$_3$As$_3$ shows some properties similar to superfluid $^3$He.

 Polycrystal sample of Na$_2$Cr$_3$As$_3$ was prepared by a low temperature ion-exchange method \cite{Na2Cr3As3}, and the others 
 were prepared by solid state reaction method \cite{K2Cr3As3}.  We crush the samples into powders or cut them into pieces to avoid skin depth problem in the NQR measurements. To protect the samples against air and water vapor, we seal the samples into an epoxy (stycast 1266) tube in an Ar-filled glove box. 
 The  \emph{T}$_1$ was measured by using the saturation-recovery method, and obtained by a good fitting \cite{yang} of the nuclear magnetization to $1-M(t)/M_{0} = \exp(-3t/T_{1})$, where $M(t)$ is the nuclear magnetization at time $t$ after the single saturation pulse and $M_0$ is the nuclear magnetization at thermal equilibrium.

\begin{figure}[htbp]
\includegraphics[width= 8.5 cm]{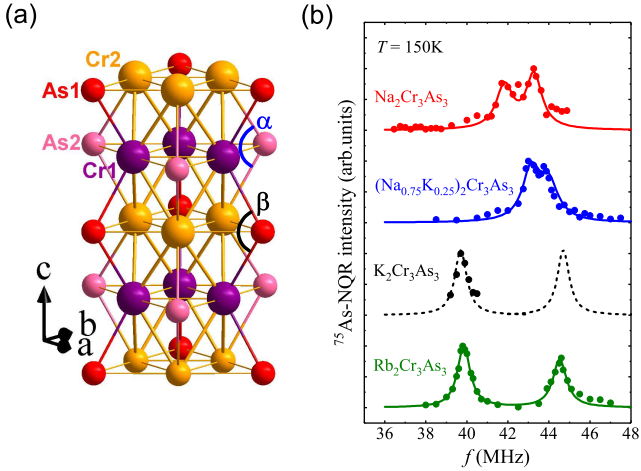}
\caption{(Color online) (a) The structure of [Cr$_3$As$_3$]$_\infty$ tube,  the   Cr2-As2-Cr2 bond angle ($\alpha$) and Cr1-As1-Cr1 bond angle ($\beta$). 
	(b) $^{75}$As NQR spectra of A$_2$Cr$_3$As$_3$(A = Na, Na$_{0.75}$K$_{0.25}$, K, Rb) measured at $T$ = 150 K. 
	 Solid lines are fitting results by two Lorentzian functions. The dashed curve represents data of   \cite{imai}.
\label{spec}}
\end{figure}
 Figure~\ref{spec}(a) and (b) show the crystal structure of the system and 
 the NQR spectra, respectively. Similar to K$_2$Cr$_3$As$_3$\cite{imai} and Rb$_2$Cr$_3$As$_3$\cite{yang}, the NQR spectra of (Na$_{0.75}$K$_{0.25}$)$_2$Cr$_3$As$_3$ and Na$_2$Cr$_3$As$_3$ also have two peaks originating from two inequivalent As sites.
 Since the temperature dependence of 1/$T_1$ for two As sites is the same \cite{imai,yang}, we measured 
 1/$T_1$ for A=Na, Na$_{0.75}$K$_{0.25}$ and K at the stronger peak.

\begin{figure}[hbp]
	\includegraphics[width= 8.5 cm]{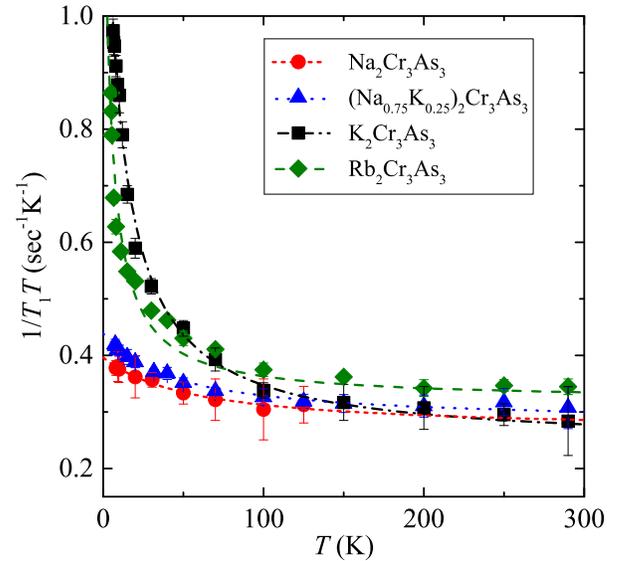}
	\caption{(Color online) $^{75}$As nuclear spin-lattice relaxation rate 1/\emph{T}$_1$ divided by temperature. 
		Data for Rb$_2$Cr$_3$As$_3$ are from Ref.\cite{yang}.
		The dashed curves on the normal state data are fittings to 1/\emph{T}$_1$\emph{T}= $a$ + $b$/(\emph{T}+ $\theta$).
		\label{T1T}}
\end{figure}

 As can been seen in Fig.~\ref{T1T}, 1/$T_1T$ is a constant above $T \sim$150 K and increases with decreasing temperature down to $T_{\rm sc}$ for all samples. For a conventional non-interacting metal, 1/$T_1T$ is a constant. Therefore, the results indicate electron correlations.
  Previous Knight shift measurements in Rb$_2$Cr$_3$As$_3$ found that spin susceptibility increases with decreasing temperature, which indicates that the electron correlation is ferromagnetic in character. 
 DFT calculations also show that the interaction  within each Cr sublattice is ferromagnetic \cite{ferromagneticfluctuation1, magnetism,pwave}. Below we apply the 3D ferromagnetic spin fluctuations theory of Moriya to characterize  the spin fluctuations \cite{SCRtheory}. The 1/$T_1T$  can be expressed as  1/$T_1T$ = (1/$T_1T$)$_{\rm SF}$ + (1/$T_1T$)$_{0}$. The first part originates from spin fluctuations of 3$d$ electrons, and the second part is due to non-interacting electrons. For  3D ferromagnetic spin fluctuations \cite{SCRtheory},  (1/$T_1T$)$_{\rm SF}$ follows a relation of  $C/(T + \theta)$, with the parameter $\theta$ describing a distance to  FM QCP. The obtained  $\theta$ 
  decreases in the order of A=Na, Na$_{0.75}$K$_{0.25}$, K, and Rb. The parameter $\theta$=4$\pm$1.5 K for Rb$_2$Cr$_3$As$_3$ and $\theta$= 57$\pm$7 K for Na$_2$Cr$_3$As$_3$.

\begin{figure}[hbp]
\includegraphics[width= 8.5 cm]{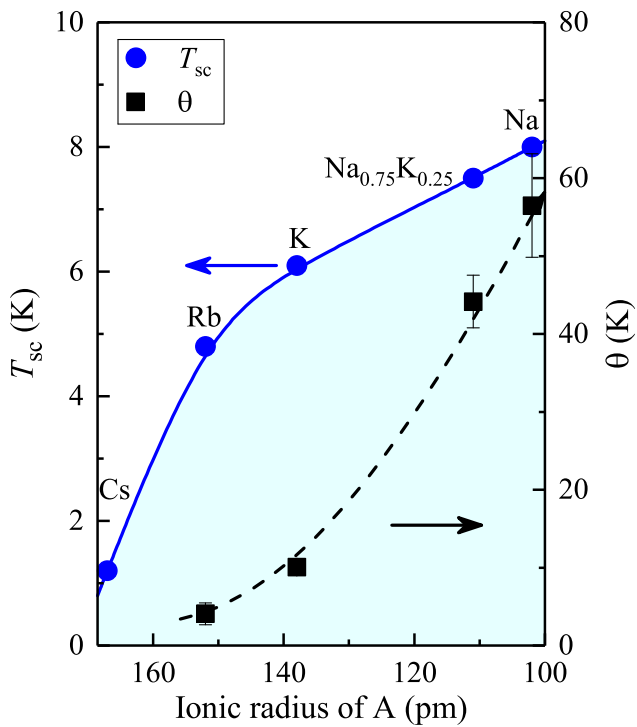}
\caption{(Color online)  Phase diagram of A$_2$Cr$_3$As$_3$(A = Na, Na$_{0.75}$K$_{0.25}$, K, Rb, Cs). 	
	The data of   Alkali ions with  six-coordinations were taken from \cite{Shannon}.
	The filled squares are $\theta$ from the fitting of  1/\emph{T}$_1$\emph{T} data (see text).
\label{phase}}
\end{figure}

Figure ~\ref{phase} shows the phase diagram of the family, where $\theta$ and $T_{\rm sc}$ are plotted as a function of the ionic radius of alkali element A. 
As the ionic radius increases, the parameter $\theta$ decreases, indicating that the system is tuned closer to a possible FM QCP. Concomitantly, $T_{\rm sc}$ decreases. Below we show that the transverse axis of Fig.~\ref{phase} correlates with ferromagnetic interaction strength. The larger the ionic radius is, the stronger the ferromagnetic interaction is.  

\begin{figure}[hbp]
\includegraphics[width= 7.5 cm]{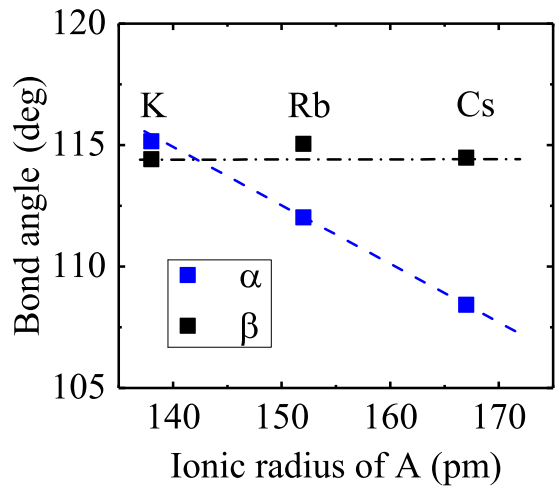}
\caption{(Color online) 
	 Correlation between the ionic radius and the Cr2-As2-Cr2 bond angle  $\alpha$ and Cr1-As1-Cr1 bond angle ($\beta$)  for   Cs$_2$Cr$_3$As$_3$, Rb$_2$Cr$_3$As$_3$ and K$_2$Cr$_3$As$_3$. The angles  were calculated based on the available crystal-structure data in the literatures \cite{K2Cr3As3,Cs2Cr3As3,233Review}.
\label{bondangle}}
\end{figure}

  As shown in Fig.~\ref{bondangle}, analysis of the available data 
  shows  that the  Cr2-As2-Cr2 bond angle  $\alpha$ has a linear relationship with the ionic radius, while the  Cr1-As1-Cr1 angle $\beta$  is almost a constant. 
Notably, as the ionic radius increases, 
 $\alpha$ decreases towards  90$^\circ$.
It was  pointed out that  the Cr-Cr coupling would be antiferromagnetic according to   Goodenough-Kanamori-Anderson rule  \cite{pwave}.
 However, since Cr is in an averaged valence of 2.3, double exchange interaction through Cr-As-Cr path is possible \cite{pwave}, which is ferromagnetic.
 At  $\alpha$=90$^\circ$, the As-4$p_x$ and As-4$p_y$ orbitals become degenerated with respect to Cr-3$d$ 
 orbitals,
 which will  maximize the double exchange interaction between 
   the two Cr2  along the $c$-axis via the As-4$p_x$ and As-4$p_y$ orbitals.
 Therefore, on going from A = Na to Na$_{0.75}$K$_{0.25}$, K, and  Rb, 
 an increase in the  ferromagnetic interaction can be expected,  
which drives the system towards 
a FM QCP. In order to directly access such QCP, we propose to replace A with Ca, Sr or Ba, thereby a long-range ordered phase can hopefully be obtained. 

In Fig.~\ref{phase}, 
one sees that  \emph{T}$_{\rm sc}$ decreases  upon approaching the FM QCP. This is in sharp contrast to the AFM case where \emph{T}$_{\rm sc}$ forms a peak around a QCP.  
Superconductivity near a FM QCP  in paramagnetic side was  discussed by Fay and Appel in their seminal work \cite{appel}, but to our knowledge, had not been confirmed thus far.
In the AFM case, the pairing interaction is enhanced due to increased quantum fluctuations \cite{Monthoux}.
In the FM case, when it is approached from the paramagnetic side,  increased quantum fluctuations also enhances pairing strength \cite{appel,Monthoux}. However, Fay and Appel found, based on a random phase approximation,
that mass enhancement due to FM spin fluctuations will kill a spin-triplet superconducting state so that
\emph{T}$_{\rm sc}$ is zero right at ferromagnetic QCP \cite{appel}.  
Later on, Monthoux and Lonzarich \cite{Monthoux} and Wang {\it et al}  \cite{WangZQ} calculated \emph{T}$_{\rm sc}$ 
in the strong-coupling limit.  
They pointed out that mass enhancement and a finite quasiparticle life time act as  pair breaking, but found a non-zero \emph{T}$_{\rm sc}$   at ferromagnetic QCP. 
 In UGe$_2$, however, no \emph{T}$_{\rm sc}$ was found  in the paramagnetic side. In a related compound UCoGe \cite{UCoGe}, although superconductivity survives after ferromagnetic order is suppressed, the ferromagnetic-paramagnetic transition is a first-order phase transition \cite{Ishida} and thus cannot be directly compared to theories.
 Our results experimentally demonstrate \cite{Supple}, for the first time,  the evolution of \emph{T}$_{\rm sc}$ in paramagnetic side predicted by theories \cite{appel,Monthoux}. Looking forward, it would be interesting to  experimentally probe the evolution of  effective electron mass by London penetration depth  at the zero-temperature limit \cite{WangCG}, for example.

\begin{figure}[hbp]
	\includegraphics[width= 8.5 cm]{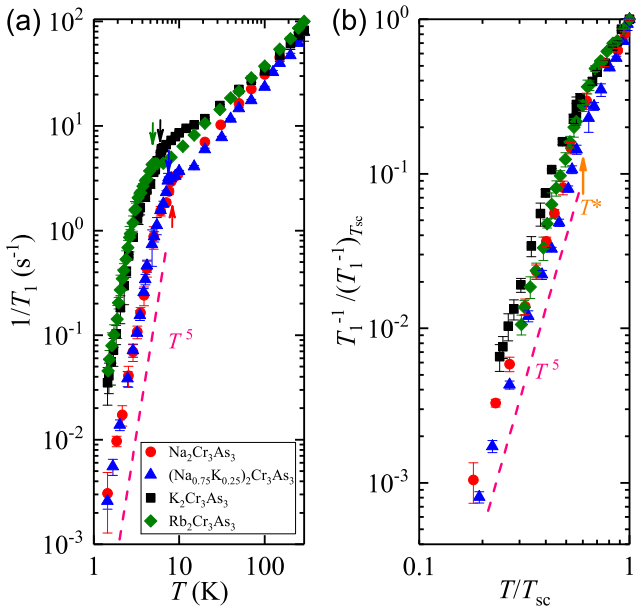}
	\caption{(Color online) (a)1/\emph{T}$_1$ as a function of $T$ for all samples. The arrows indicate \emph{T}$_{\rm sc}$ of each compound. (b)1/\emph{T}$_1$ normalized by its value at \emph{T}$_{\rm sc}$. The arrow indicates a characteristic temperature \emph{T}*, below which 1/\emph{T}$_1$ becomes proportional to \emph{T}$^5$. The symbols are the same as (a). 
		\label{T1}}
\end{figure}

Next, we discuss the properties of  superconducting state. In Fig~\ref{T1}(a) are shown the 1/\emph{T}$_1$ data for the four compounds.  There is no coherence peak for all cases. 
In contrast to the previous report for a K$_2$Cr$_3$As$_3$ sample with a lower $T_{sc}$=5.7 K than our case (6.1 K) \cite{imai,Supple}, our data show that  K$_2$Cr$_3$As$_3$ exhibits a  temperature variation very similar to Rb$_2$Cr$_3$As$_3$.
 In Fig~\ref{T1}(b), we show the data with reduced scales for the axes. 
 As can be seen  there, 1/\emph{T}$_1$ commonly shows a $T^5$ behavior below a characteristic $T$*/\emph{T}$_{\rm sc}$ $\approx$ 0.6. This  suggests that the gap symmetry is the same for all members of this family.

The  1/\emph{T}$_1$ in superconducting state can be expressed as
\begin{equation}
\frac{(T_1)_{T_{\rm sc}}}{T_{\rm 1}} = \frac{2}{k_BT}\int N_{\rm s}(E)^2(1+\frac{\triangle^2}{E^2})f(E)(1-f(E))dE,
\end{equation}
where $N_{\rm s}(E)$ is the DOS below \emph{T}$_{\rm sc}$, $f$(E) is Fermi distribution function, $\Delta$ is the gap function.
When there exist point nodes in the gap function, such as in the
ABM model \cite{ABM1,ABM2} with  $\Delta =\Delta_0sin\theta e^{i\phi}$, $N_{\rm s}(E)\propto E^2$. This
results in 1/\emph{T}$_1$ $\propto$ $T^5$ at low temperatures.
Previously, Katayama $et$ $al$  found such a $T^5$ variation in filled skutterudite superconductor PrOs$_{\rm 4}$Sb$_{\rm 12}$ under pressure \cite{T5}.
 Notably, 1/\emph{T}$_1$ does not deviate from $T^5$ in the lowest temperature even for (Na$_{0.75}$K$_{0.25}$)$_2$Cr$_3$As$_3$ where  partial substitution of K for Na would induce disorder. 
  Generally, disorders or impurities cause finite DOS at the Fermi level in the case of nodal gap, which will result in a deviation of 1/\emph{T}$_1$ 
  , which  has indeed been observed 
  in cuprates \cite{ingapstate}. The current result  therefore indicates that the superconducting state is not affected by the disorder out of [Cr$_3$As$_3$]$_\infty$ tube.
  Finally, the existence of $T$* is unclear at the moment. A possible reason 
  is multiple-bands superconductivity, as seen in iron-based superconductors \cite{Matano}. 
  Another possibility is multiple phases arising from internal freedoms associated with  spin-triplet pairing as seen in UPt$_3$ \cite{multiphase}. 
  Clearly, more work is  needed in this regard.

In summary, we have performed $^{75}$As-
NQR measurements on the   A$_2$Cr$_3$As$_3$ family. 
we find that 
by changing A in the order of A=Na, Na$_{0.75}$K$_{0.25}$, K, and Rb,
the system is tuned to approach a possible ferromagnetic QCP.
We propose that the Cr2-As2-Cr2 bond angle  ($\alpha$) that  decreases  towards 90 degree is responsible for the  increase of ferromagnetic interaction. 
Upon approaching 
the QCP, the superconducting critical temperature $T_{\rm sc}$ decreases,
which is in sharp contrast to the AFM case where $T_{\rm sc}$ usually forms a broad peak around  a QCP.
In the superconducting state,
 1/$T_1$ decreases   with no Hebel-Slichter peak just below $T_{\rm sc}$, and ubiquitously follows a $T$$^{5}$ variation at low temperatures, 
which indicates the existence of point nodes in the  gap function commonly in the whole family.
Our results indicate  that the A$_2$Cr$_3$As$_3$ family is a possible solid-state analog of superfluid $^3$He. Therefore, further investigations on this family promise to enrich the physics across multiple research areas of electron correlations, unconventional superconductivity and topological quantum phenomena.
We thank G.H. Cao, Y. Haga, J.P. Hu, H. Kontani, K. Miyake, A. Sumiyama, Z.Q. Wang, H.M. Weng, G.M. Zhang    and Y. Zhou for helpful discussions, and C.G. Wang for technical assistance. 
This work was partially supported by NSFC  grants (Nos. 11674377,  11634015 and 11774399), MOST grants (Nos. 2016YFA0300502, 2017YFA0302901 and  2017YFA0302904), 
  as well as JSPS/MEXT Grants (No. JP15H05852 and JP19H00657). J. Y. also acknowledges  support by the Youth Innovation Promotion Association of CAS.


\begin{thebibliography}{0}
\bibitem{Luojununpublished}
We have also performed NQR measurements on Cs$_2$Cr$_3$As$_3$ down to $T$=1.4 K and obtained similar results with Ref. \cite{imaiCsCrAs}. Namely, different from other compounds described in this paper, $1/T_1T$ for Cs$_2$Cr$_3$As$_3$ shows a  pseudogap-like behavior below $T\sim$75 K, whose origin is unknown. 
\end{thebibliography}

\end{document}